# Transthoracic super-resolution ultrasound localisation microscopy of myocardial vasculature in patients

Jipeng Yan[1], Biao Huang[1], Johanna Tonko[2], Matthieu Toulemonde[1], Joseph Hansen-Shearer[1], Qingyuan Tan[1], Kai Riemer[1], Konstantinos Ntagiantas[3], Rasheda A Chowdhury[3], Pier Lambiase[2], Roxy Senior[3], Meng-Xing Tang[1*]

Author information:
[1]Ultrasound Lab for Imaging and Sensing, Department of Bioengineering, Imperial College London, London, UK.
[2]Institute of Cardiovascular Science, University College London, London, UK.
[3]Faculty of Medicine, National Heart & Lung Institute, Imperial College London, London, UK.
These authors contributed equally: Jipeng Yan, Biao Huang.
*Corresponding author: Meng-Xing Tang (mengxing.tang@imperial.ac.uk)

## Abstract

Micro-vascular flow in the myocardium is of significant importance clinically but remains poorly understood. Up to 25% of patients with symptoms of coronary heart diseases have no obstructive coronary arteries and have suspected microvascular diseases. However, such microvasculature is difficult to image in vivo with existing modalities due to the lack of resolution and sensitivity. Here, we demonstrate the feasibility of transthoracic super-resolution ultrasound localisation microscopy (SRUS/ULM) of myocardial microvasculature and hemodynamics in a large animal model and in patients, using a cardiac phased array probe with a customised data acquisition and processing pipeline. A multi-level motion correction strategy was proposed. A tracking framework incorporating multiple features and automatic parameter initialisations was developed to reconstruct microcirculation. In two patients with impaired myocardial function, we have generated SRUS images of myocardial vascular structure and flow with a resolution that is beyond the wave-diffraction limit (half a wavelength), using data acquired within a breath hold. Myocardial SRUS/ULM has potential to improve the understanding of myocardial microcirculation and the management of patients with cardiac microvascular diseases.

## Introduction

Myocardial micro-vasculature is of significant and increasing importance clinically. For example, for Coronary Heart Disease (CHD), which is the leading cause of mortality worldwide, Computer Tomography Coronary Angiography (CTCA) is the recommended first line investigation by the most recent NICE guidelines[1] in patients with suspected angina or those who are asymptomatic with suggested ECG changes for ischemia. However, CTCA assessment is limited to only large artery obstruction, despite that between 40-50%[2,3] of patients with symptoms of suspected CHD have ischemia with non-obstructive coronary arteries (INOCA). This population includes a heterogenous group of patients with various etiologies including coronary artery vasospasm and importantly coronary microvascular disease (CMD), with the latter being evident in more than half of the patients with INOCA[4]. Comparing to large artery disease, our understanding of CMD is still very limited. Initially thought to be a combination of structural and functional changes at the level of the microvasculature, recent studies have led to a more differentiated understanding and sub-classification to structural CMD and functional CMD subtypes[5,6]. They may both manifest with stress perfusion defects on cardiac Magnetic Resonance Imaging (MRI) yet differ on a pathophysiological level and hence accurate characterisation may have important clinical implications for patient management and choice of therapeutic interventions. CMD is also associated with non-ischemic cardiomyopathy, inflammatory cardiac disease and even with non-obstructive coronary plaque disease.

Existing clinical modalities have challenges in imaging the micro-vascular structure and flow in the myocardium in vivo. Myocardial microvasculature refers to vessels within the heart wall and are typically hundreds of microns or less in diameter[7]. Imaging is not only challenging due to their small size but also their constant movement throughout the cardiac cycle. Both clinical CT, MRI and nuclear imaging methods lack the spatial resolution and/or contrast to directly visualise these small vessels and the hemodynamics in them. Recently ultrafast ultrasound[8,9] has shown promise in imaging such myocardial vessels *ex vivo* and *in vivo*. However, the blood cell only generates weak scattering signal limiting its signal to noise ratio (SNR), and its image resolution is restricted by the classical wave diffraction limit.

Super-resolution ultrasound (SRUS), also known as ultrasound localisation microscopy (ULM), can image deep microvasculature with resolution beyond the diffraction limit by accurately localising microbubbles (MBs) from the contrast-enhanced ultrasound (CEUS) images[10–14]. Hemodynamics in the microvasculature can be measured in SRUS/ULM by tracking the movements of super-localised MBs[15–18]. SRUS has been used to image vasculature in many human organs, including lower limbs[19], breast[20], liver[21] and brain[22]. The concept of using SRUS for imaging coronary vasculature in myocardium has been recently demonstrated in small animals in 2D and 3D [23,24].

However, clinical translation of SRUS for non-invasive imaging of myocardial microvasculature remains a challenge, mainly due to the large cardiac motion, the required large field of view and significant penetration depth. SRUS imaging requires detection and accumulation of MB localisations over time, assuming the vascular structure remaining stationary[13,14]. During the accumulation any tissue motion, inevitable for *in vivo* imaging, would require correction to not degrade the image. Motion correction is a cross-cutting theme in medical imaging technology, and in SRUS some efforts have been made to correct such motion in e.g. lower limbs[19], rat heart [23,24], and rat kidney[25], although in cardiac imaging it is particularly challenging due to the large and complex cardiac motion. Furthermore, unlike the pre-clinical imaging where a high frequency probe (central frequency typically between 6-15 MHz) with an aperture size similar to that of the animal heart is used[23]. Clinical imaging would require a low frequency probe (typically 1-4 MHz) to achieve appropriate penetration depth. Additionally, as the human heart is larger and due to challenges in imaging through the ribs, the clinical aperture size will be much smaller than the heart. Such low frequency, small aperture, combined with diverging waves required for ultrafast acquisition, usually generate lower SNR and spatial resolution in deep tissue, posing additional challenges for detection and tracking of MBs.

**Results**

This work presents the first demonstration of transthoracic myocardial SRUS/ULM in human patients with reconstructed super-resolution vasculature and blood flow dynamics in the myocardium, using a customised data acquisition and processing pipeline (Fig. 1). In the pipeline, a pulse sequence is designed for low frequency ultrasound (2 MHz) to detect and enhance CEUS signals with MB motion correction among steering angles; coherence to variance (CV) beamformer is incorporated to achieve high SNR and reduce side lobe artefacts; a two-level tissue motion correction strategy is proposed to deal with the inter- and intra-cardiac cycle motions; a feature-motion-model framework is developed to track MBs. The feasibility of the pipeline was demonstrated on an *ex vivo* Langendorff porcine heart and on two patients with impaired left ventricular function.

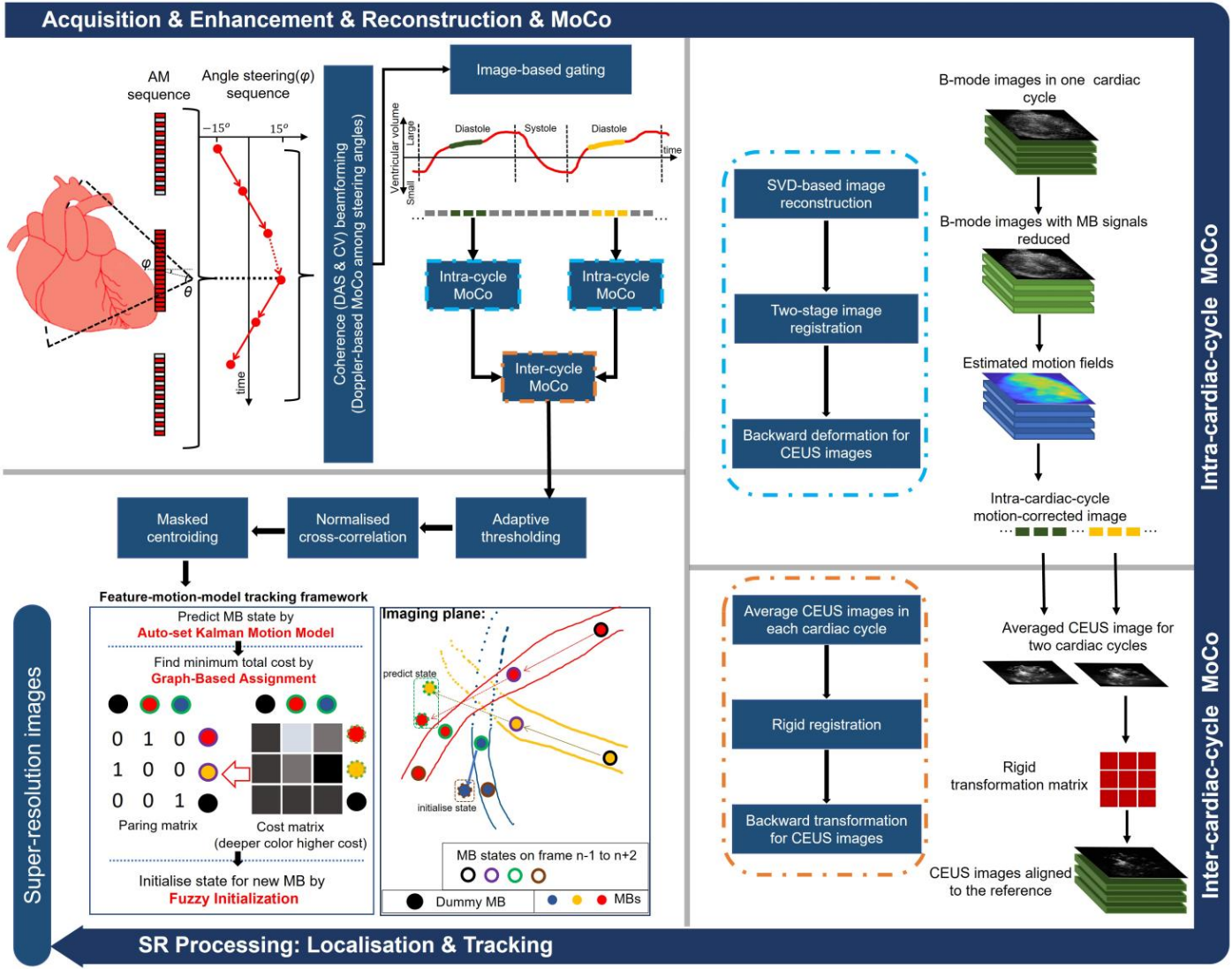

**Fig. 1|** *In vivo* data acquisition and processing pipeline. A three-pulse AM sequence was used to detect and separate tissue and CEUS signals. Diverging waves were steered in six angles in a triangle sequence for angle compounding. MB movement between steering angles was detected and corrected to effectively enhance the contrast signal after angle compounding. Frames in diastole of the cardiac cycle were excluded by using a frame intensity correlation-based gating algorithm. A two-level, intra-cardiac-cycle and inter-cardiac-cycle, image registration strategy was proposed to correct complex myocardium motions. MBs in the motion-corrected CEUS images were localised with normalised cross-correlation and paired by a proposed feature-motion-model tracking framework. Super-resolution images were plotted by accumulating MB trajectories. *Ex vivo* data processing had the same pipeline without gating and inter-cardiac-cycle motion correction (MoCo).

A clinical cardiac phase array probe (M5Sc-D, GE HealthCare, NY, USA) was driven by a research ultrasound system (Vantage 256, Verasonics, WA, USA) at a central frequency of around 2 MHz. Diverging waves were transmitted at an ultrafast pulse repetition frequency of 5490 Hz. Amplitude modulation (AM) was used for separating MB and tissue signals. By steering diverging waves in six angles, a frame rate of 305 Hz was achieved after compounding.

Sonovue (Bracco, Milan, Italy) MBs were used as ultrasound contrast agents in this study. For the *ex vivo* porcine heart imaging, MBs were infused by a syringe pump (Harvard Apparatus, Holliston, MA, USA) at an infusion rate of 5 ml/min. For the *in vivo* human heart imaging, 2 ml of MBs were manually injected in a slow bolus (around 6 seconds). In this study, 5 seconds and 10 seconds datasets were acquired on *ex vivo* and *in vivo* hearts, respectively. For *ex vivo* heart scanning, we used both parasternal short- and long-axis view. For *in vivo* hearts, we also acquired on parasternal short- and long-axis views, but the available views depended on the acoustic window of patients.

To separate MB signals from tissue signals in the radiofrequency (RF) channel data, AM pulse sequence was firstly used, and the remaining tissue signals were further reduced by subtraction of moving average across three frames. The B-mode and CEUS images were reconstructed in polar coordinates by Delay and Sum (DAS) beamformer and Coherence to Variance (CV) beamformer respectively. A Doppler-based motion estimation[26] method was used to correct the movement of MBs among steering angles and enhance the contrast signals after angle compounding. Evidence of the improvement by the Doppler-based MB motion correction can be seen in Fig. S1 of the supplementary figures.

A two-stage, affine and B-splined-based nonrigid, image registration approach[19] was used to compensate myocardium motions in the porcine heart. For the *in vivo* human dataset specifically, an image-intensity-based gating algorithm was firstly implemented to select and index frames in the diastolic phase, which were with least motions among the whole cardiac cycle. The remaining motion of the myocardium within the diastolic phase in each cardiac cycle was corrected using the two-stage image registration approach after removal of the MB signals from the tissue signals by Singular Value Decomposition (SVD) filtering.  Finally rigid image registration was used to align CEUS images across different cardiac cycles. Effectiveness of this two-level strategy is demonstrated when comparing SR density maps obtained with and without motion correction, as shown in Fig. S2 of the supplementary figures.

For each CEUS image, image patches containing bubble signals were segmented out by binary pixel connectivity after adaptive thresholding. Normalised cross-correlation was applied to each MB patch and the corresponding spatially varying Point Spread Function (PSF) of the region where the MB signals are located. Single MB images were cropped out of the image patches with the mask based on the cross-correlation coefficient map and the MB was localised through centroiding. The MBs were tracked to generate the flow dynamic information for the microvasculature and enhance the saturation of reconstructed SRUS images with the MB trajectories. A feature-motion-model framework proposed in our previous work[27] was improved to track MBs. Firstly, fuzzy initialisation of the motion model was proposed to estimate the initial velocity of newly appeared MB to further improve the Kalman motion model performance. Secondly, a method to estimate the Kalman filter parameters from the data was also proposed to make the framework free from manual adjustment. Details of the tracking framework can be found in the supplementary method. SR images were plotted by accumulating the MB trajectories in the map, each track having a width defined by a 2D Gaussian whose full width at half maximum (FWHM) was set as a quarter of a wavelength. Since the MB concentrations were too high for localisation and tracking individual bubbles inside the aortic, ventricle, and atrium, those regions were cropped out in the analysis.

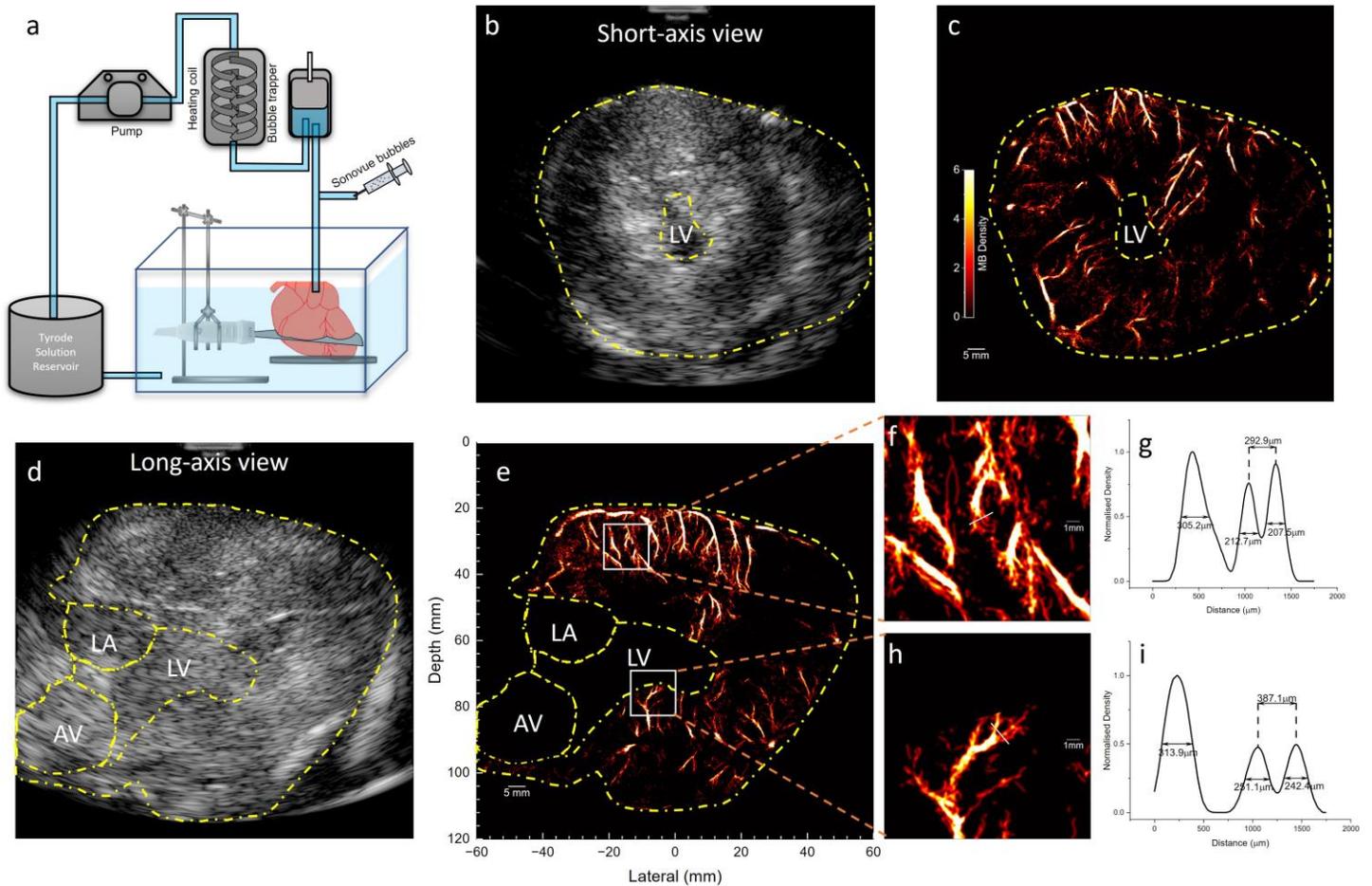

**Fig. 2|** Langendorff *ex vivo* porcine heart data acquisition and results. **a**, the Langendorff setup for *ex vivo* dataset acquisition. **b, d,** the B-Mode images from short- and long-axis views of the heart and the yellow dashed lines delineated the region cropped out in the processing. **c, e,** the corresponding SR density map for the microvasculature inside the myocardium. **f, h**, zoomed in density map of white boxes in **e** and white solid lines cut the vessel for cross-section analysis. **g, i,** the normalised density profile from the cross-section analysis.

In the *ex vivo* porcine Langendorff heart experiment, the probe's position was adjusted and clamp-fixed to have a good field of view for the short-axis and long-axis of the heart (Fig. 2a). From the B-Mode images of short-axis view (Fig. 2b) and long-axis view (Fig. 2d), myocardium surround the left ventricle (LV), left atrium (LA) and aortic valve (AV), delineated with dashed lines, can be visualised. The corresponding ULM MB density maps reconstructed from 5-seconds acquisitions were present alongside of the B-Mode images (Fig. 2c and 2e). Detailed microvasculature inside the left ventricle wall and the ventricular septum was presented in the density maps. MB density profiles along the white dotted lines in the magnified regions (Fig. 2f and 2h) were measured and plotted after being normalised by the maximum respectively (Fig. 2g and 2i). Vessels separated by distances of 292.3 µm and 387.1 µm can be distinguished from the profiles. The vessels' size was estimated by the FWHM and ranged from 207.5 µm to 313.9 µm.

In the clinical acquisition, the probe was manually positioned by a clinician to acquire parasternal short- and long-axis views in the patients. Positions of the imaging plane are shown in Fig. 3b. Both views were acquired for the patient one but only a short-axis view for the patient two was acquired, as the long-axis view was limited by the acoustic window of the patient two. After gating, a total of 1110 frames (include 10 cardiac cycles), 990 frames (include 9 cardiac cycles), 630 frames (include 9 cardiac cycles) within the diastolic phase were used to reconstruct SRUS images in the short- and long-axis views of the patient one and the short-axis view of the patient two respectively.

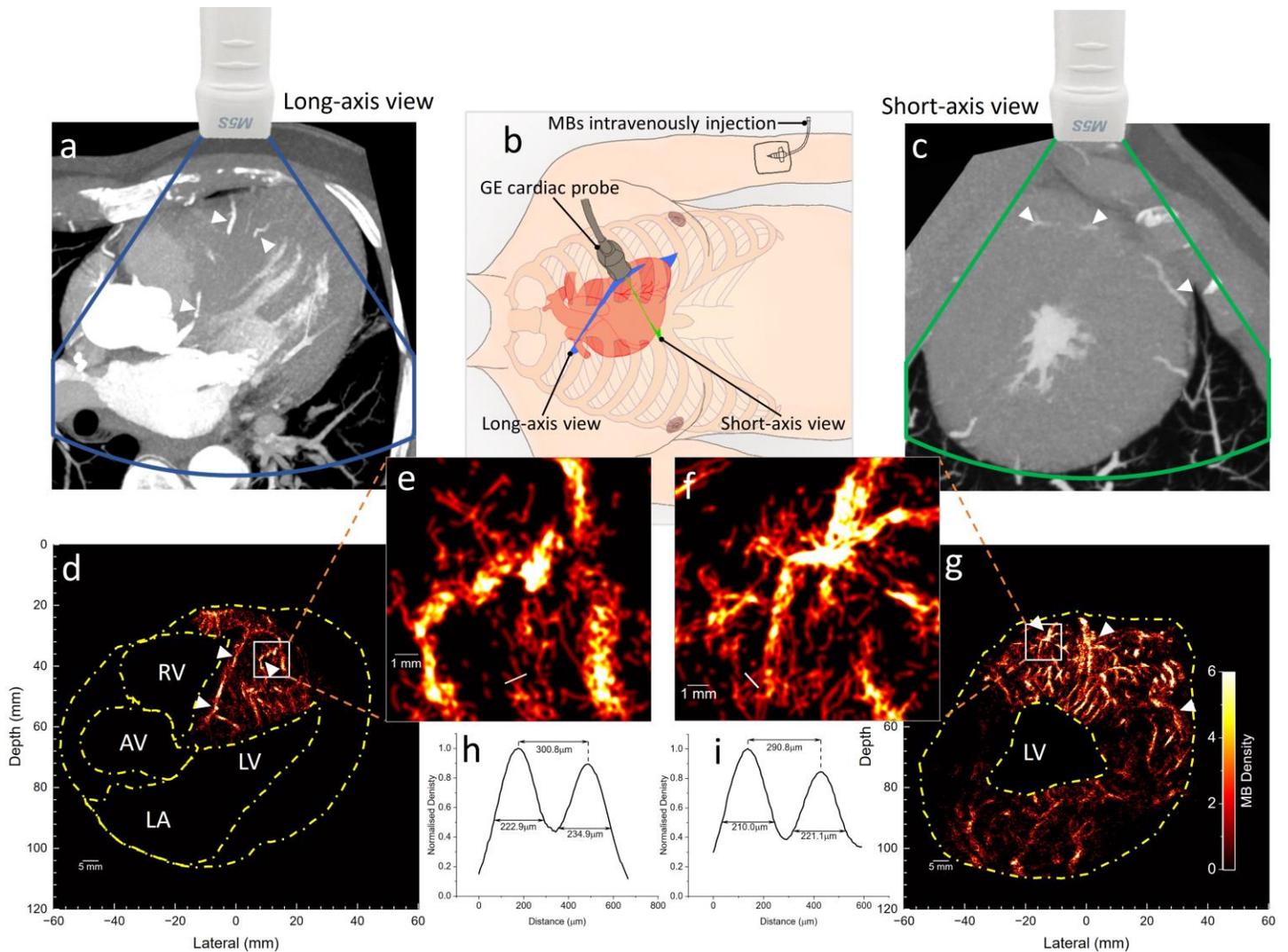

**Fig. 3|** *In vivo* CTCA scans and SRUS imaging for patient one. **a, c,** the long- and short-axis views in the CTCA scans approximated equivalent plane as those in the ultrasound scans. White arrows point out the similar structures that are visible in both modalities. **b,** positions of the ultrasound probe on the chest for cardiac imaging. Blue and green planes indicate the long- and short-axis views of the heart respectively. **d, g,** myocardial SRUS density maps. The yellow dashed lines indicate the chamber regions that was cropped out. **e, f,** zoomed-in density map of white boxes in **d** and **g**, white solid lines cut the vessel for cross-section analysis. **h, i,** the normalised density profile from the cross-section analysis. **Overlapped images of two modalities can be found in the supplementary video.**

Microvasculature inside the interventricular septum and the left ventricular wall of first patient is visible from the ULM MB density maps in both the long- and short-axis views (Fig. 3d and 3g). The closest equivalent views taken from multiplane reconstructions of the corresponding diastole CTCA scans are shown in Fig. 3a and 3c. Compared with the clinical diastole CTCA images, SRUS exhibits increased sensitivity and resolution for myocardial microvessels. Some similar large vessels appear to be in both CTCA and SRUS images (Fig.3 white arrows), although it should be noted that the CT images correspond to a slab with a constant thickness of 10 mm, while the SRUS has a much smaller slab thickness (a typical cardiac probe has a slab thickness of a few wavelength). Many small vessels, from the epicardium penetrating the middle wall to the endocardium are visible in the short-axis view (Fig. 3g), while in the CT scan only several major vessels were detected. The cross-section analysis was also performed on two visually separated vessels in the magnified regions (Fig. 3e and 3f). As shown in Fig. 3h and 3i, these two pairs of vessels are 300.8 μm and 290.8 μm apart, compared to half the wavelength of 320 μm. MB density map for patient two (Fig. 4b) also presents detailed microvasculature in the myocardium. Compared to the equivalent view of CT scan (Fig. 4a), two big vessels are also visible on the SRUS results (white arrows in Fig. 4), but many vessels not visible in CT become clearly visible on SRUS results, as shown in the zoomed-in microvasculature density map (Fig. 4c, 4d).

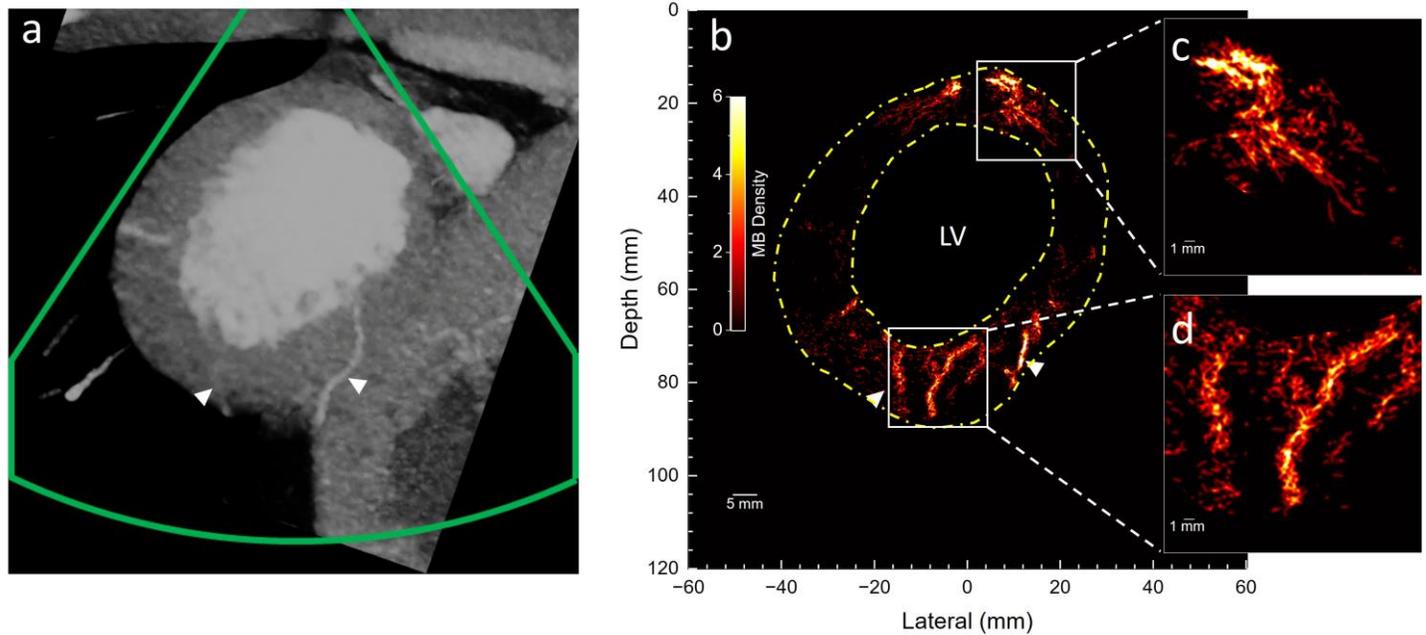

**Fig. 4|** *In vivo* CTCA scan and SRUS images for patient two. **a,** the short-axis view in the CTCA scan that corresponds to the ultrasound scan plane. White arrows point out the similar structures that are visible in both modalities. **b,** the SRUS density map for myocardial microvasculature. **c, d**, the zoomed-in SRUS density map for the region in **b**.

Image resolution was also estimated by a Fourier ring correlation (FRC)-based method through randomly dividing MB trajectories in two subsets[28]. Based on the FRC estimation with ½ bit information threshold, the imaging resolution for *ex vivo* long- and short-axis views were 58.2 and 57.2 µm, the resolution for patient one in long- and short-axis views were 58.8 and 57.8 µm, and the resolution for short-axis view of patient two was 59.2 µm. The above estimated resolutions are all more than 5 times higher than the half wavelength of the transmission pulses (320 µm for 2.4 MHz and 452 µm for 1.7 MHz).

Hemodynamics in the myocardium can be presented by the flow speed maps (Fig. 5a and 5d), which were obtained by tracking MB movements across frames in SRUS processing, not available in conventional CTCA scans. The hemodynamics can also be presented in flow direction maps (Fig. 5b and 5c), where the flow between epicardium and endocardium can be visualised. Animations of the reconstructed data with MB flowing inside microvasculature can be found in the supplementary video. Two quantitative metrics, vessel diameter and flow speed distributions, in the short-axis view of the two patients were calculated (Fig. 6). There appear to be more smaller vessels in patient one who was with hypertrophic cardiomyopathy (HCM) than patient two who was without obvious HCM, which is consistent with a previous finding by histological analysis that narrowed arterioles were in HCM patients[29]. The mean flow speed in the two patient's coronary vessels were measured to be 54.3±23.9 vs. 52.7±24.3 mm/s.

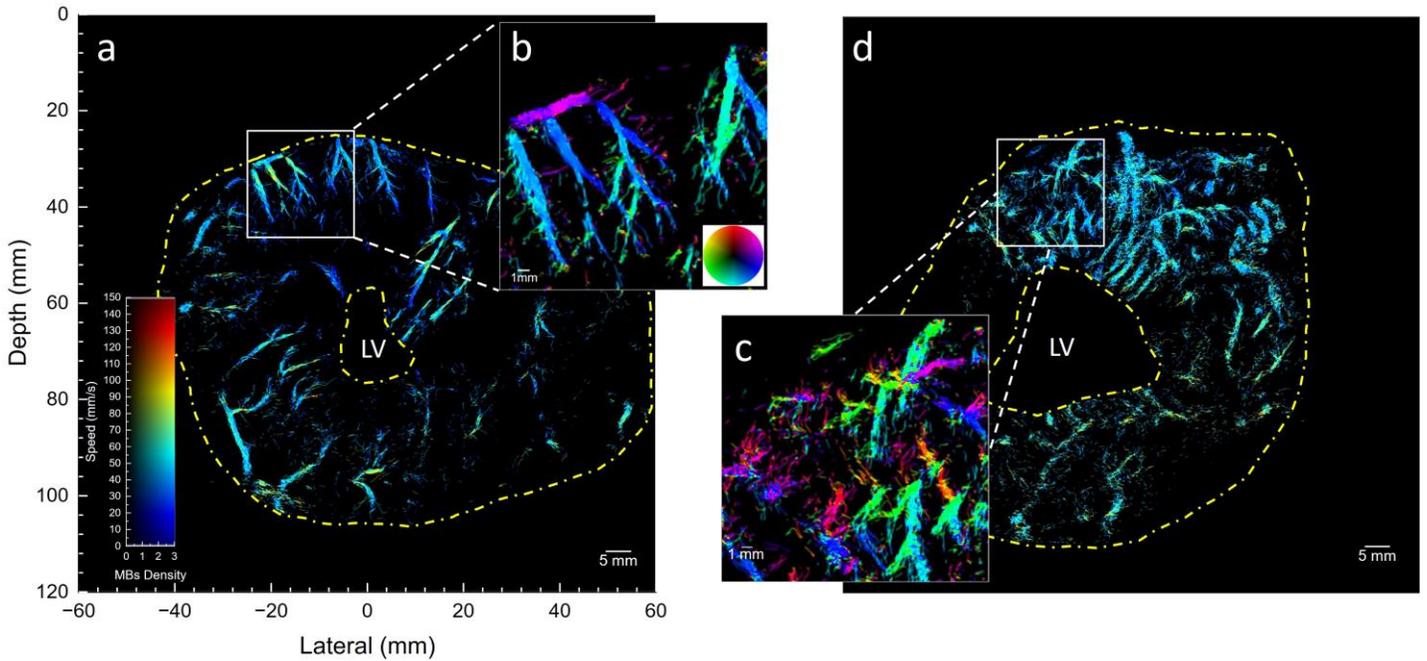

**Fig. 5|** *Ex vivo* porcine heart and *in vivo* patient short-axis views of flow speed and direction maps. **a, d**, the SR flow speed maps, corresponding to the SR density maps in **Fig. 2** and **3**. **b, c,** SR flow direction maps of zoomed-in regions inside the white boxes in **a** and **d**. The speed and direction colormaps are shown in **a** and **b**.

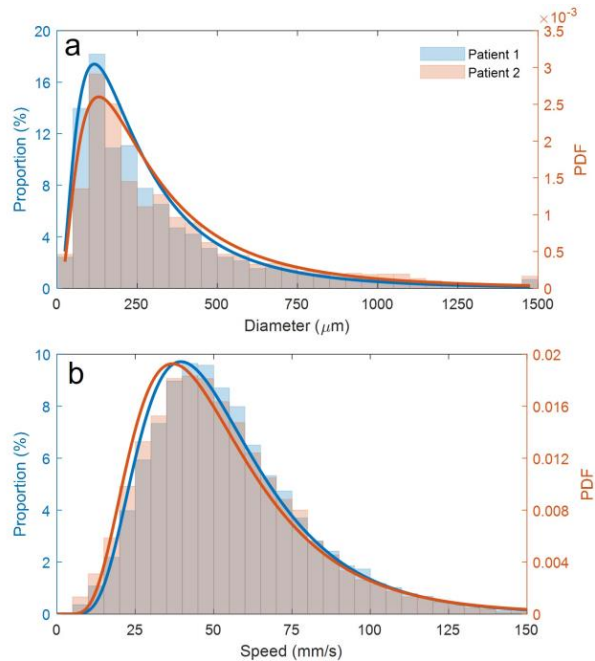

**Fig. 6|** Quantitative metrics obtained from SR images of short-axis view of the two patients. **a**, normalised distribution of vessel diameter. Bin width is 50 μm. Vessel diameter larger than 1500 μm are counted as 1500 μm. Bin height is calculated by the ratio of the length of vessels with diameter in the corresponding bin range to the total vessel length. **b**, normalised speed distribution in the flow speed maps of the short-axis view of the two patients. Bin width is 5 mm/s. Bin height is calculated by the ratio of the number of pixels with value in the corresponding bin range to the total number of pixels with non-zero values. Lines are created by fitting corresponding data by log normal distribution which has been used to describe the flow velocities in lymph node vessels[30]. PDF: probability density function.

## Discussion

To our knowledge, this study demonstrates for the first time the clinical feasibility of transthoracic SRUS of myocardial microvascular structure and flow, using a customised data acquisition and processing pipeline. Compared with CTCA, SRUS shows more detailed myocardial microvasculature at sub-diffraction resolution, can be accessed at bedside and is more affordable and free from ionising radiation. The high resolution and sensitivity to myocardial vascular flow and its quantification potentially leads to better understanding of microcirculations in the myocardium and improved management of patients with cardiac microvascular diseases.

The clinical SRUS acquisitions were undertaken in a similar setup as that for echocardiograph examination in a cardiac outpatient clinic using an existing clinical echocardiography transducer, facilitating the clinical translation of the technique. The employed CE-marked contrast agent, Sonovue, was first approved in the European Union more than 20 years ago and is currently used in 36 countries. The intravenous bolus injection of the contrast agents is an established clinical operation for clinical contrast imaging examination. The acquisition for SRUS was done with a single ultrasound imaging system on a trolley, which is suitable for bedside investigation. One main limitation of the SRUS technique is that currently it is processed offline due to computational power of the hardware and computational speed of the algorithm, which can be optimised in the future.

While SRUS has revealed both macro- and micro-vessels in the myocardium in the two patients, these are only a small sample of the large vascular network in the myocardium, and there are regions with no apparent vessels identified. In order to reveal more of the myocardial vasculature, there are important factors to consider:

Firstly, the data acquisition time and number of CEUS image frames are very low in this case. The slice of heart in the imaging plane can change significantly due to breathing and heart twisting motions. In order to reduce the effect large and out-of-plane motion with 2D ultrasound, we asked the patients to hold breath and only used the frames during diastoles for generating the SRUS image. Therefore, number of frames available to us for cardiac SRUS imaging is limited by the maximum time for patients to hold breath and the available frames in the diastoles. In this case the total acquisition time is up to 10 seconds, out of which only around 3 seconds of data during diastoles was used for reconstructing the final SRUS image. This is orders of magnitude less than those typically used e.g., in brain ULM[22,31] where no breath holding and cardiac gating are required, and contribute to the limited saturation of vasculature in the reconstructed cardiac SRUS images. 3D imaging could help address this issue, which is discussed later.

Secondly, the sensitivity and specificity of detecting individual MBs with sufficient SNR is important and can be further improved. Compared to existing pre-clinical studies on small animals, the SNR of acquired clinical CEUS images is more challenging due to the significant signal attenuation of the diverging waves over depth. Furthermore, there was still tissue breakthrough in the AM image that reduces the sensitivity of individual MB detection. When moving average background subtraction was used to remove the tissue breakthrough, signals from very slow-moving MBs might be removed together with tissue signals. The sensitivity, specificity and SNR of the MB images could be improved by e.g. using nonlinear Doppler[32]. Compared with other organs, the SNR and bubble specificity decrease further when ultrasound passes through the cardiac chambers filled with MBs causing attenuation, phase aberration, and nonlinear propagation. This may explain the darker SRUS images in the lower part of the myocardium in Fig. 3g.

Future work to develop ultrafast 3D cardiac ultrasound can facilitate motion correction and further improve the myocardial SRUS. Correction of 3D cardiac motion is feasible if 3D + time imaging can be achieved using e.g. a matrix array transducer. This could potentially enable a much longer data acquisition while patients have shallow breathing. Furthermore, imaging frames within a longer duration of a cardiac cycle could be used if 3D motion is corrected. Therefore, SRUS images reconstructed with 3D imaging can be expected to achieve significantly higher vascular saturation. 3D imaging also reduces the need to accurately position the transducer and improve the volumetric coverage of the imaged object. Covering a larger volume could make the quantification of the myocardial vascular flow more robust.

## Methods

1. **Porcine heart/Patient management**

For the *ex vivo* experiment, a porcine heart was explanted from a large white female pig (65-75kg/ 4-5 months old). The heart was extracted as previously described[33] under anaesthesia using a human donor heart retrieval protocol.

Animal studies were reviewed by the Royal Veterinary College Animal and Ethical Review Board and carried out in accordance with ethical standards (European Commission 2010, the Animal Welfare Act 2006 and the Welfare for Farm Animals (England) Regulations 2007).

The patient one was a 60-year-old patient (male, BMI 24 kg/m2 – 167 cm, 67 kg) with a typical hypertrophic cardiomyopathy and secondary prophylactic left-sided dual chamber implantable cardioverter defibrillator (ICD). The patient two was a 42-year-old patient (female, BMI 23 kg/m2 – 170 cm, 66 kg) with a non-dilated mildly impaired left ventricular function without evidence of myocardial scar in the cardiac MRI. The left ventricular function was likely secondary to frequent monomorphic ventricular ectopy with a burden of 36%.

Both patients were recruited from the cardiac outpatient clinic after giving informed and written consent to participate in a prospective clinical cohort study investigating novel methods for non-invasive arrhythmogenic substrate characterisation including of high frame rate contrast-enhanced transthoracic echocardiography. The study was reviewed and approved by the London-Bromley Research Ethics Committee and the Health Research Authority (IRAS Project ID 144257, REC reference 14/LO/0360) and is ongoing.

2. **Acquisition**

2.1 **Coronary CT angiography**

The prospective ECG-gated CT coronary angiography scans were performed using a Siemens SOMATOM scanner. Following administration of sublingual GTN and intravenous metoprolol, axial cross section of the heart was acquired with patients in the supine position upon inspiratory breath hold.

2.2 **Ultrasound sequence**

The acquisition of ultrasound images for *ex vivo* and *in vivo* heart was performed using a Vantage Verasonics acquisition system and a phased array ultrasonic probe (GE M5ScD, GE; central frequency, 2.84MHz; 90% bandwidth at -6dB; pitch, 0.27mm; and 80 x 3 elements, one row in centre and two 80 half-height elements at sides). The ultrafast ultrasound virtual source of the diverging wave was set at 21.6 mm above the centre of the probe, giving a 53° angular field of view. The imaging sequences consisted of the emission of 6 angles of diverging waves, in ascending (-15°, -3°, 9°) and descending (15°, 3°, -9°) triangle sequence[26], which was designed for Doppler-based motion estimation. Sonovue MBs were used as the ultrasound contrast agents. The gas-encapsulated MBs compress and expand asymmetrically when exposed to ultrasound waves. Such nonlinear echo-response can be exploited for separating MB signals from tissue signals with the Amplitude Modulation (AM) sequence in each steering angle, using a full, half and half interleaved aperture approach. The probe was fired at the maximum pulse repetition frequency (5490 Hz) allowed by the desired imaging depth of 120 mm. With angle compounding and coded AM transmission, the frame rate was 5490÷6÷3 = 305 Hz.

Ultrasound was transmitted to achieve a mechanical index (MI) of 0.05-0.07 at the depth of 30 mm when using the full aperture. 5-second and 10-second acquisitions were made on the *ex vivo* heart and on patients, respectively. For each acquisition, backscattered echoes were received by the phase array probe and digitalised by Verasonics system at four times of the transmitted centre frequency. RF data of each channel were stored for off-line image reconstruction and processing in MATLAB (R2021a, MathWorks, WA, USA) on a desktop (CPU: AMD Ryzen 9 5900 Processor, GPU: Nvidia GeForce RTX3080, RAM: 128 GB).

3. **Image reconstruction**

The echoes of the three pulses in the AM sequence were combined in each channel to generate the contrast specific AM signals. Likely due to nonlinear propagation[34], tissue signals could not be completely cancelled out by the AM sequence. To reduce the residual tissue signals, a temporal moving average subtraction was applied to RF channel data across three frames for each steering angle based on the assumption that tissue remains static but MBs move within the temporal averaging window, which is about 10 ms.

Hilbert transform was used to obtain analytic channel signals. B-mode images were reconstructed by the Delay-and-Sum beamforming method with coherent angle compounding[35], keeping the linearity of image intensity to meet the requirement of the Single Value Decomposition (SVD) that was used in tissue motion correction. Contrast-enhanced ultrasound (CEUS) images were reconstructed by the Coherence to Variance (CV) beamforming method[36] to reduce the side lobes and noise, benefitting the MB isolation for super-localisation. The CV beamformer calculates an adaptive weight for each pixel by the ratio of the squared coherent sum to the variance across all the channels and all the steering angles. The pixel value is then generated by the multiplication of the adaptive weight and the coherent sum.

Multi-angle compounding assumes stationary MBs between the steering angles, while in reality MBs move and hence the compounded MBs signals decorrelate[37]. In this study, a Doppler-based method[26] was used to estimate MB velocities among the steering angles and then the CV beamformer was implemented after correcting MB signal shifts. Note that the phase rotator used in literature[26] was discarded when correcting the signal shifts, as Hilbert transform, instead of In-phase and Quadrature (IQ) demodulation, was used here. Images were reconstructed in polar coordinates for MB motion correction, envelope-detected and transferred in Cartesian coordinates with spline interpolation for tissue motion correction and tracking in SRUS processing.

**4. Image-based tissue motion correction**

*In vivo* tissue motions were corrected at two levels, intra-cardiac-cycle and inter-cardiac-cycle, considering the complexity of heart beats and limitation of 2D ultrasound. While out-of-plane motion correction was challenging due to loss of signals, we focused on the diastole, where twisting motion is the least among all the phases in a cardiac cycle.

We applied an image intensity-based gating algorithm to the dataset. The cardiac cycle presents a periodically change of pixel intensity distribution between frames, which can be tracked from change of the normalised cross-correlation coefficient between two adjacent frames. The frame with a lowest correlation coefficient appears during one cardiac cycle indicates the start of systole. By automatically finding different phases of a cardiac cycle, we then only take data during the diastoles, between 0.2 to 0.4 seconds of data, for further analysis and SRUS image generation.

To correct the intra-cardiac-cycle tissue motions, a two-stage, affine and then B-spline-based nonrigid[38], image registration was conducted on the B-mode images to estimate and correct tissue motions[19]. Reference image for motion correction was obtained by finding the frame with maximal similarity, measured by MATLAB *'ssim'* function, to the average frame of this cycle. Sum of squared error was used to quantify the image difference between moving and reference image. B-mode ultrasound images in one cardiac cycle were processed by SVD and reconstructed with the 5% largest singular values to reduce the effect of moving MB signals on motion estimation. Reconstructed B-mode images were log-compressed with a dynamic range of 50 dB and rescaled to grey scale. The two-stage image registration began from the reference frame and transformation matrices estimated on the current frame were set as the starting point for the next frame, reducing the computation time and improving temporal smoothness of the motion fields.

Following the correction of intra-cardiac motion, the inter-cardiac-cycle motions were subsequently corrected through a rigid transformation using CEUS images which have clear anatomical signals from chambers and large vessels useful for registration. Firstly, CEUS images were averaged in each cardiac cycle to reduce the image difference due to flowing MBs, and rigid transformation then was estimated between the averaged CEUS images to correct motions between cardiac cycles. The averaged CEUS images were also log-compressed and rescaled to grey scale before registration. Note that while out-of-plane motion cannot be ruled out, they would bring additional information from neighbouring slices of the myocardium.

For the *ex vivo* dataset, the tissue motions were corrected by applying the two-stage image registration to all the acquired frames, as there is no breathing, probe motion or heavy twisting motion of the heart during the acquisition.

**5. Super-resolution processing**

Adaptive thresholds were obtained by MATLAB 'adaptthresh' function for each frame. Pixels with intensity lower than the adaptive thresholds or the estimated noise level were set to zero to reduce noise and side lobes in the CEUS images. MBs patched in each frame were generated using the connectivity of non-zero pixels. Considering the spatially varying PSFs in images generated using diverging-wave, the image was divided into 5 (depth) x 5 (lateral) regions and PSF in each region was estimated by averaging 10 single MB images that were segmented from the CEUS sequence. Normalised cross-correlation[17] was calculated between each patch of MBs and the PSF in the region where the patch centre was. Single MB images were cropped out of the patches by the mask of cross-correlation coefficient map whose value was over 0.5. MBs were localised by the pixel-intensity-weighted centroid of each MB image and round to a pixel map with resolution of 13.5 µm, and the intensity of an MB was given by the sum of pixel values in the corresponding single MB image. MBs were tracked in a feature-motion-model framework[27]. MBs in two frames were paired by finding the global minimum of costs in the graph-based assignment[39]. The cost of a candidate MB pair was defined as the ratio of the image intensity difference normalised by the stronger intensity of the two paired MBs to the probability obtained by a linear Kalman motion model[15,40,41] established with positions of MBs. An MB could be paired to a dummy MB if its pairing costs to any other MBs were over a limit or there were different numbers of MBs in the two frames. The MB paired to the dummy MB was regarded as either a disappeared or newly appeared MB. Additionally, fuzzy initialisation method was proposed to initialise the Kalman state vector of a newly appeared MB with estimated velocity, instead of assuming the MB to be static initially, to further improve the performance of the motion model. A parameter estimation method was proposed to automatically estimate the covariance of model prediction noise to avoid human interaction in setting up the motion model and to avoid the effort of tuning the parameter. Explanations of MB tracking problem with Shannon theory and detailed descriptions of the tracking framework can be found in the supplementary method.

6. **Display (SR maps and animation)**

MBs tracked over no less than four frames were kept for plotting images. Trajectories of flowing bubbles were plotted by linking paired MB positions by straight lines. The localisation density map was plotted by accumulating trajectories with a width defined by a 2D Gaussian, using MATLAB function 'imgaussfilt', whose FWHM was set as a quarter of ultrasound wavelength at the transmitted centre frequency for visualisation.

To plot the flow speed map, MB moving speeds were accumulated on each pixel of the trajectory. Accumulated speed and accumulated localisation density maps were both filtered by a disk whose diameter is same as the Gaussian FWHM. Flow speed map was calculated by division between the two disk-smoothed maps. To draw the flow direction map, accumulated MB velocities were smoothed by the same disk and then divided by the disk-smoothed density map. The flow direction on each pixel was calculated from the averaged velocity direction. Flow dynamics maps were presented by HSV colours with hue to illustrate speed or direction and value to illustrate MB density.

Animated flow was rendered by moving each localised bubble position along its trajectory with the estimated speed. Warm or cold colour indicates MB moving upwards or downwards.

7. **Parameter calculation**

The centre lines of vessels were detected on the binarized density map by MATLAB "bwmorph" function. The diameters of vessels were defined by double the distance between vessel centre lines and boundaries. The vessel distribution corresponding to a range of diameters was counted with the pixels on centre lines. Distribution of blood flow speeds were calculated from the non-zero pixel values on the SR speed maps.

## Data availability

Sample dataset is available on the figshare with the identifier (https://doi.org/10.6084/m9.figshare.22300903). All raw and analysed ultrasound data used in this study are available on reasonable request.

## Code availability

SRUS software, containing user guide interface and codes developed by us, is provided on GitHub (https://github.com/JipengYan1995/SRUSSoftware). Estimated PSFs, localisation uncertainty, maximum blood flow speed for searching MBs used for processing the data used in this study were set as default parameter in Version 2.1. Sample dataset for testing localisation and tracking of the software is provided on figshare with the identifier (https://doi.org/10.6084/m9.figshare.22300903).

## Acknowledgements

We thank the Royal Veterinary College and Prof. Justin Perkins for assistance with the porcine heart extraction. We thank Dimitrios Panagopoulos, Danya Agha-Jaffar, Shengzhe Li and Clara Rodrigo Gonzalez for the help with the *ex vivo* heart experiment. This work was supported by the Chan Zuckerberg Foundation under Grant No. 2020-225443, the Engineering and Physical Sciences Research Council under Grant No. EP/T008970/1 and EP/R511547/1, Rosetrees Trust Grant M645 to RAC, China Scholarship Council, the Wellcome Trust under Grant No. 222845/Z/21/Z, the National Institute for Health Research i4i under Grant NIHR200972.

## Author contributions

- Meng-Xing Tang conceived the study.
- Rasheda A Chowdhury and Konstantinos Ntagiantas designed the Langendorff *ex vivo* setup.
- Meng-Xing Tang, Biao Huang, Matthieu Toulemonde, Johanna Tonko and Jipeng Yan designed the *in vivo* setup.
- Matthieu Toulemonde and Biao Huang developed the acquisition sequence.
- Biao Huang, Konstantinos Ntagiantas, Matthieu Toulemonde, Kai Riemer and Qingyuan Tan acquired *ex vivo* data.
- Biao Huang, Johanna Tonko and Joseph Hansen-Shearer acquired *in vivo* data.
- Meng-Xing Tang and Jipeng Yan designed the data processing pipeline.
- Matthieu Toulemonde and Jipeng Yan developed the image reconstruction algorithm.
- Biao Huang and Jipeng Yan developed motion correction algorithm.
- Jipeng Yan and Meng-Xing Tang developed the super-resolution imaging framework and software.
- Biao Huang and Jipeng Yan processed the data.

- Biao Huang, Johanna Tonko, Jipeng Yan, Meng-Xing Tang, Pier Lambiase and Roxy Senior interpreted the results.
- Jipeng Yan, Biao Huang and Meng-Xing Tang wrote the first draft.
- All authors edited and approved the final version of the manuscript.

## Competing interests

## Supplementary Figures

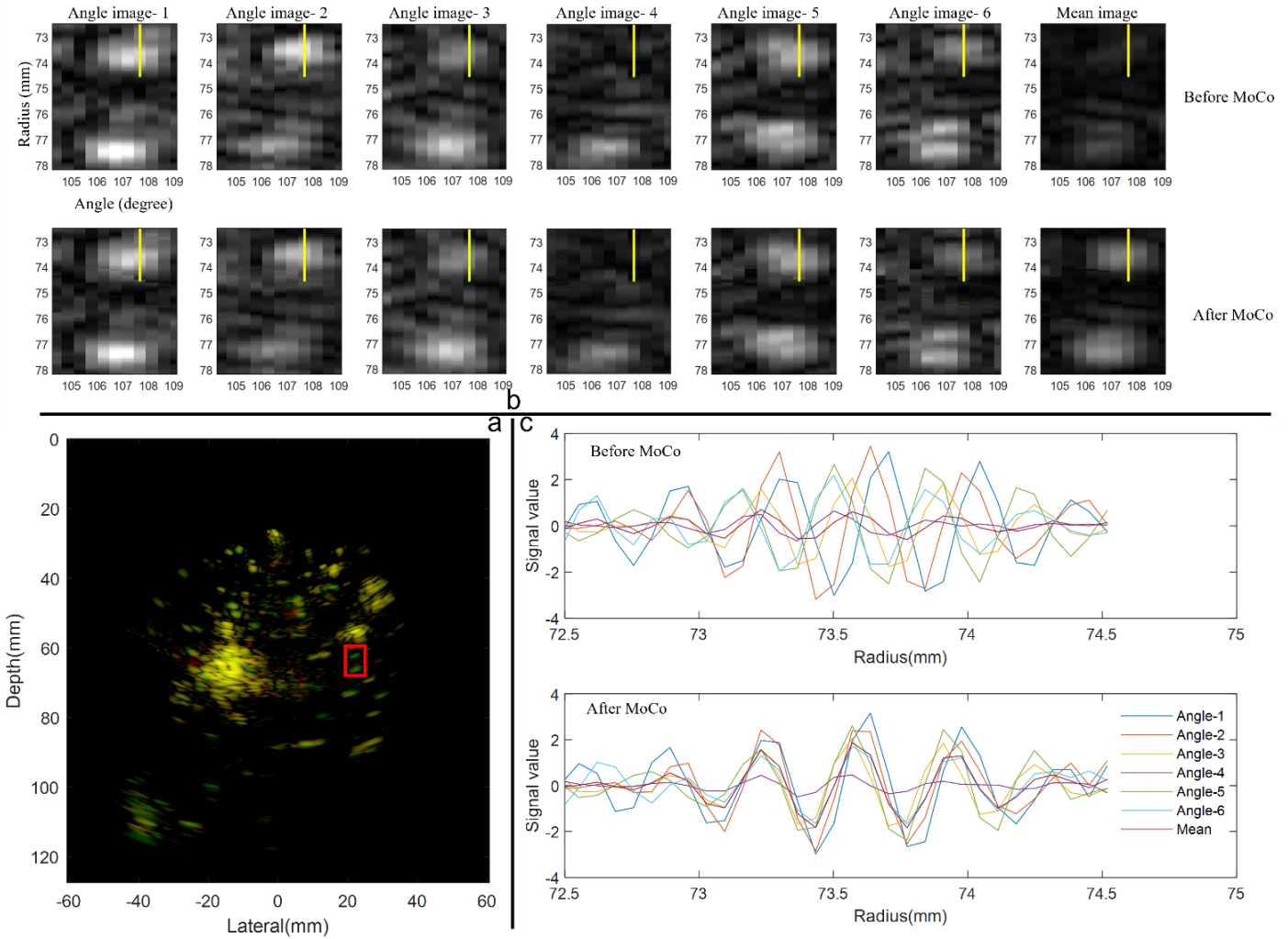

**Fig. S1|** Evidence of improvement by using Doppler-based MoCo for compounding, noted as Angle Moco for convenience. **a,** compounded images reconstructed by DAS with and without Angle MoCo were normalised, log compressed with dynamic range of 40dB, and then overlayed by Matlab 'imfuse' function, using red for the image without Angle MoCo, green for image with Angle Moco, and yellow for areas of similar intensity between the two images. Green colour cover more area, at which there seems to be single bubbles. **b,** magnified image of the region in red box in **a**. Images were plotted in polar coordinates, linear scale and same dynamic range ([0, 4]), after envelope detection. Compounded image obtained by averaging angle image with Angle MoCo presents stronger intensity than its counterpart. **c,** real part of the signals along the yellow line in **b**. Signal shifts among steering angles can be corrected and thus gives a higher average.

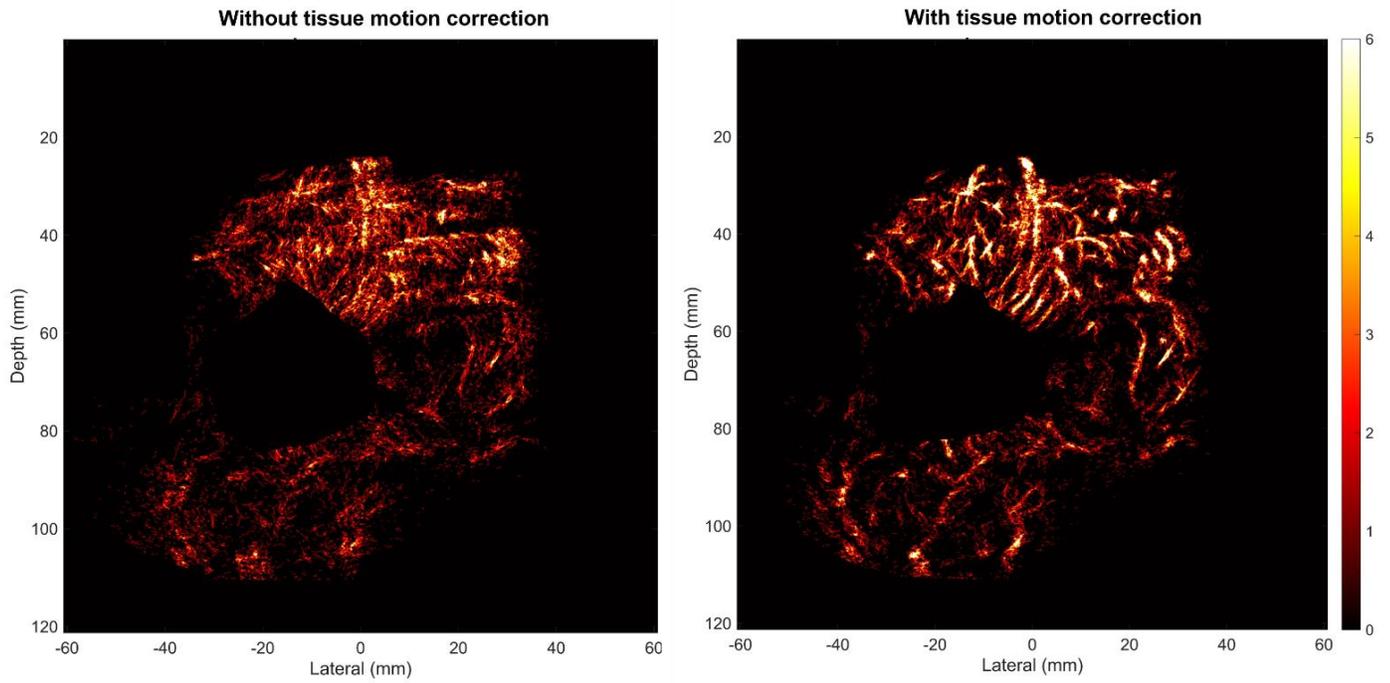

**Fig. S2|** Motion effect on the short-axis view acquisition of the patient in Fig. 3. The comparison between the density map reconstructed without and with tissue motion correction presents obviously visual difference.

# Supplementary Method: Fuzzy Initialisation and Parameter Estimation for Tracking based on Kalman Motion Model

In this supplement, we explain the impacts of microbubble (MB) concentrations and ultrasound acquisition frame rates on MB tracking, giving the reasons for developing the methods and unifying concentrations and frame rates as one equivalent factor in tracking problem; present the method for Kalman state vector initialisation; present the method for Kalman parameter estimation.

1. **Shannon theory in MB tracking**

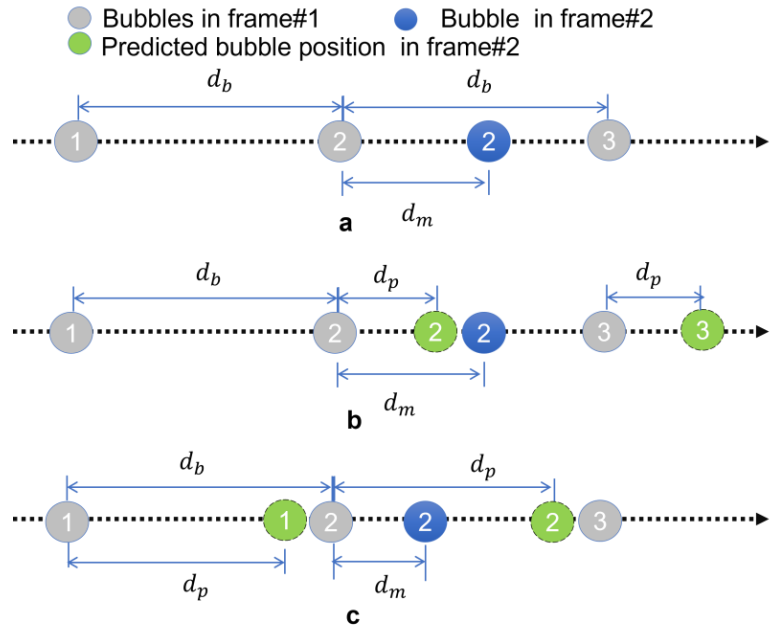

**Fig. S3|** Multiple MBs moves in one straight vessel where MB 2 in frame 2 will be paired with MBs in frame 1. **a,** a case where MBs are paired by nearest-neighbour-based method. **b,** a case where MBs are paired by motion model with predicted movement less than the moving distance. **c,** a case where MBs are paired by motion model with predicted movement over the moving distance. Number in each disk denotes the identity of an MB. $d_b$ is the distance between MBs in one frame. $d_m$ is the moving distance of the MB between two frames. $d_p$ is the moving distance predicted by motion model for MBs.

The relation between Shannon theory and the tracking can be revealed by simple cases, where multiple MBs move in a straight vessel, as shown in Fig. S3a. To rightly pair MB 2 by the nearest-neighbour method, the distance between the positions of MB 2 in two frames should be smaller than the distance between MB 2 in the second frame and MB 3 in the first frame.

$$d_m < d_b - d_m \quad i.e. \quad d_m < 0.5 d_b \qquad (1)$$

where $d_m$ is the moving distance, the ratio of flow speed to frame rate, and $d_b$ is the MB distance which is inverse to the MB concentration. According to the above equation, a higher frame rate is needed to track MBs at higher concentrations and faster speeds. Motion model can predict the MB position along the flow direction, as shown in Fig. S3b. To rightly pair MB 2 in this case, the distance between the MB 2's predicted position and actual positions should be smaller than the distance between the MB 3's predicted position and MB 2's actual position,

$$d_m - d_p < d_b + d_p - d_m \quad i.e. \quad d_m < 0.5 d_b + d_p \qquad (2)$$

where $d_p$ is the MB moving distance predicted by a motion model. With the motion model, tracking method can rightly pair faster MBs at the same frame rate and MB concentrations, compared to the nearest-neighbour method. However, it doesn't mean the larger the $d_p$ is, the higher accuracy the tracking can get. To avoid pairing MB 2 in Frame 2 to MB 1 in Frame 1 in the case shown in Fig. S3c, $d_p$ should satisfy

$$d_p - d_m < d_b - d_p + d_m \quad i.e. \quad d_p < 0.5d_b + d_m \tag{3}$$

Combining Eq. (2) and (3), we can obtain

$$|d_m - d_p| < 0.5d_b \tag{4}$$

With the frame rate $f$, the equation can be rewritten as

$$\frac{2|v_m - v_p|}{d_b} < f \tag{5}$$

where $v_m$ is the flow velocity and $v_p$ is the predicted velocity and $v_m/d_b$ can be regarded as the spatial repetition frequency, defined by how many times an MB passes through the positions of the other MBs in one second. Then, the temporal frequency, frame rate or pulse repetition frequency $f$, should be at least double the spatial repetition frequency to rightly pair the MB by nearest-neighbour. Using the motion model, the spatial repetition frequency of the MB movement $v_m/d_b$ can be reduced by $v_p/d_b$, which can be analogised as IQ demodulation in ultrasound signal processing.

Fast-moving MBs at high concentrations can be more accurately tracked using motion model than only using the nearest-neighbour, especially when the frame rates are restricted by physical imaging depth and the number of angles used for compounding.

2. **Fuzzy initialization**

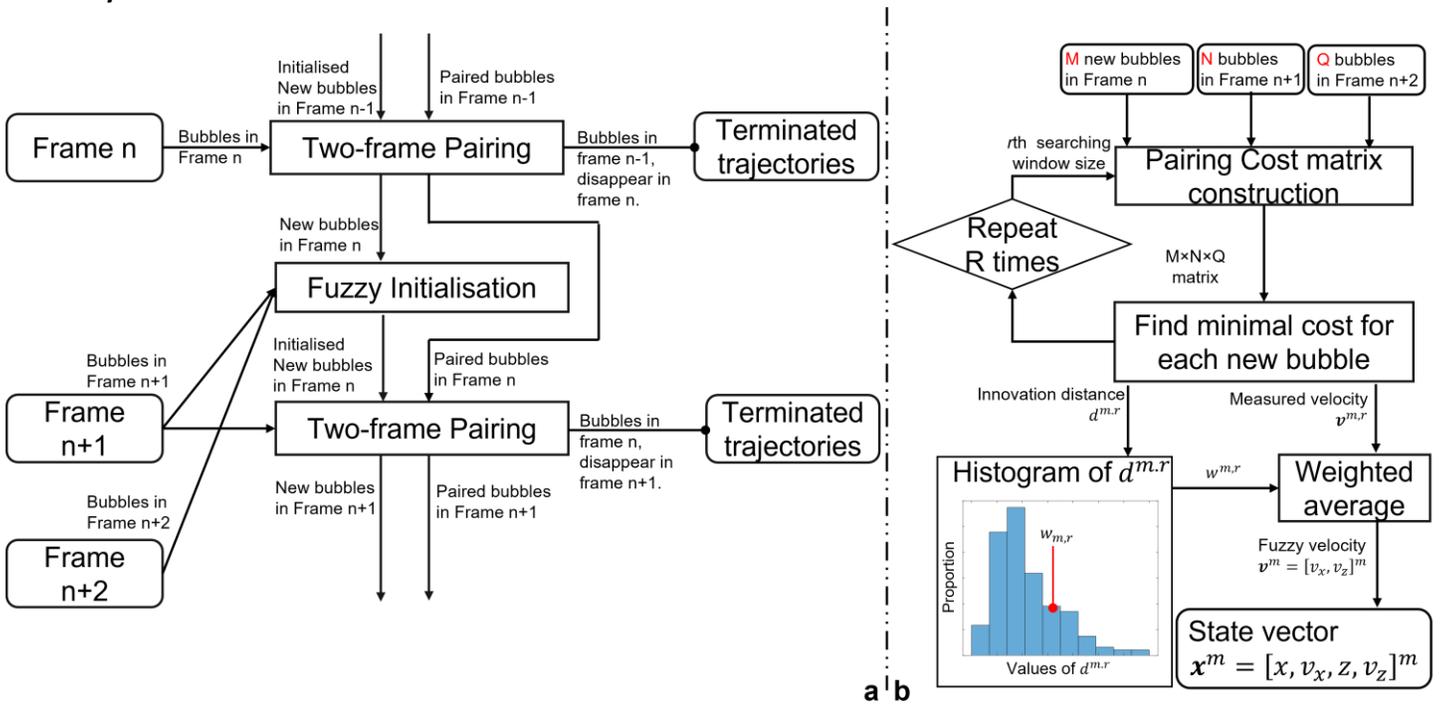

**Fig.S4|** **a,** a framework combining fuzzy initialisation and graph-based assignment. Fuzzy initialisation is done for MBs detected to be newly appeared in frame n by using MBsin frame n, n+1 and n+2. After initialisation, all MBs in frame n are paired with MBs in frame n+1 by our previously proposed feature-motion-model tracking framework. **b,** fuzzy initialisation diagram. Moving velocities of a newly appeared MB is estimated with MBs in next two consecutive frames by finding minimum of pairing cost under different searching window sizes. Velocities obtained under different searching window sizes are weighted averaged as the velocity components in the Kalman state vector of the new MB. $v^{m,r}$ is measured velocity between the first two frames corresponding to the minimal cost found for the $m$th new MB under the $r$th searching window size. Innovation distance $d^{m,r}$ in the third frame corresponds to the minimal cost found for the $m$th new MB under the $r$th searching window size.

In our previous work [1], we proposed a feature-motion-model framework to achieve accurate and efficient MB tracking, which combined MB image feature and Kalman motion model in a cost function and paired MBs by finding the total minimum cost in a graph-based assignment. The Kalman motion model's state vector **x**, translation matrix **F** and observation matrix **H** is defined as

$$\mathbf{F} = \begin{bmatrix} 1 & \frac{1}{f} & 0 & 0 \\ 0 & 1 & 0 & 0 \\ 0 & 0 & 1 & \frac{1}{f} \\ 0 & 0 & 0 & 1 \end{bmatrix}, \quad \mathbf{x} = \begin{bmatrix} x \\ v_x \\ z \\ v_z \end{bmatrix}, \quad \mathbf{H} = \begin{bmatrix} 1 & 0 & 0 & 0 \\ 0 & 0 & 1 & 0 \end{bmatrix} \quad (6)$$

where $x$ and $z$ are the lateral and depth location of MBs in the image; $v_x$ and $v_z$ are the MB moving speeds along the corresponding directions. $v_x$ and $v_z$, i.e. $v_p$, were set to 0 for the newly appeared MBs as we could not get the MB velocity before pairing. As demonstrated in the last section, wrong pairing is still likely to happen for new MB with this initialisation where motion model does not take effect by assuming new MB to be static. Therefore, we proposed fuzzy initialisation and combined it with our previous framework in the way shown in Fig. S4a. The fuzzy initialisation was done for the newly appeared MBs using three consecutive frames, as shown in Fig. S4b. The fuzzy initalisation was done in below steps.

1) **Cost matrix construction**. To initialise the $M$ new MBs in the first frame, a 3D cost matrix for paring the $M$ new MBs with the MBs in the following two frames was built. Various information can be used for calculating the cost. In this study, the cost consisted of three components: innovation of motion model, intensity differences, and intensity variances along the candidate tracks.

We estimated $v_x$ and $v_z$ for each new MB in the first two frames, and therefore the innovation of motion model could be calculated by the distance between the observed MB locations and the predicted MB locations in third frame

$$d^{m,n,q} = \sqrt{(x^q + x^m - 2x^n)^2 + (z^q + z^m - 2z^n)^2} \quad (7)$$

where the superscripts, $m \leq M$, $n \leq N$ and $q \leq Q$, denote the MB indexes in the three frames. When giving a search window, model innovations for pairs out of the searching window were given as infinity.

The intensity difference $d_I$ among MBs in three frames was calculated as

$$d_I^{m,n,q} = |I^q + I^m - 2I^n| \quad (8)$$

where $I$ is the MB image intensity.

The maximum intensity projection (MIP) of contrast enhanced ultrasound (CEUS) sequence present the vasculature in a low resolution. While pairing two MBs, the trajectory can be along or across two vessels. It can be assumed that pixel intensities on the trajectory along one vessel have smaller variance than those on trajectories across vessels. Therefore, the intensity variances of MIP pixels on each trajectory can be used for calculating the cost of pairing and defined as

$$\sigma_p^{m,n,q} = \frac{\sigma_{pt}}{\mu_{pt}} \quad (9)$$

where the numerator is the standard deviation of pixel intensities on a trajectory and the denominator is the average. To save computation, trajectory between one pair was defined as straight line; seven points were equidistantly sampled on the trajectory between MB pairs in the first two frames and then assigned to candidate pairs among three frames which shared the same pairing between the first two frames; the pixel intensity of the points on each trajectory was obtained from the nearest MIP pixel.

Three components were in the same size of 3D matrix and combined to a 3D matrix by the first-order principal components of the principal components analysis (PCA).

2) **Finding the minimum.** The minimal cost was found for each new MB independently, instead of finding the total minimum of all the pairs to save the computation. The innovation distance $d^m$ and velocity vector $v^m$ between MBs in the first two frames corresponding to the minimal cost were saved for each new MB.

3) **Repetition within different sizes of searching windows**. It is not trustable enough to get the right pair with the local minima. Considering blood flow might vary from zero to the maximum in different vessels in the imaging plane, above two steps were repeated for different search window sizes. In this study, the R, set as 10, searching window sizes differed from 0.3 to 1.2 of the MB moving distance corresponding to desired maximal blood flow velocity, 150 mm/s. As a result, for the $r^{th}$ searching window, $d^{m,r}$ and $v^{m,r}$ were saved for the $m$th new MB.

4) **Initialising Kalman state vector**. An intuitive way to combine R of $v^{m,r}$ into one for the new MB is the average. However, some extreme cases, such as too long or short vectors, among the R vectors might significantly affect the average. As the distance $d^{m,r}$ can be regarded as errors of the linear motion model, the normalised histogram, shown in Fig. S4b, represents the probability of the certain value of errors when assuming linear MB movements for this data. Therefore, the velocity of the new MB can be calculated by the below weighted average to reduce the effect of extreme cases

$$\boldsymbol{v}^m = \frac{\sum_{r=1}^{R} w^{m,r} \boldsymbol{v}^{m,r}}{\sum_{r=1}^{R} w^{m,r}} \tag{10}$$

where $w^{m,r}$ is the number obtained by indexing the histogram by $d^{m,r}$ corresponding to the $\boldsymbol{v}^m$. A new bubble might not be paired with any other bubble in a small searching window. In this case, the searching window was excluded from the average in Eq. (10). Only the new bubbles were paired in more than two searching windows were initialised by Eq. (10), and otherwise were set as static.

In the above initialisation, none of the pairing in each searching window was absolutely trusted and the speed components in the state vector were obtained by the weighted average. As a result, a velocity vector from an initialised MB might not point to any MB in the next frame. Therefore, we named the above procedures as fuzzy initialisation.

After being initialised, new MBs in current frame and MBs existing in previous as well as current frames were paired with all the MBs in the next frame by finding the total minimum of cost through the graph-based assignment. The cost for the graph-based assignment was defined as the ratio of MB image intensity difference to the probability obtained through motion model by $p(o_k|x_{k|k-1})$

$$p(\boldsymbol{o}_k|\boldsymbol{x}_{k|k-1}) = \frac{1}{2\pi|\mathbf{S}_k|^{0.5}} \exp\left(-0.5(\boldsymbol{o}_k - \boldsymbol{\mu})^{\mathrm{T}} \mathbf{S}_k^{-1}(\boldsymbol{o}_k - \boldsymbol{\mu})\right)$$
$$\boldsymbol{\mu} = \mathbf{H}\hat{\boldsymbol{x}}_{k|k-1} \tag{11}$$

where the subscript $n|n'$ denotes the estimate of the value in frame $n$ given observations up to and including frame $n'$; $o_k$ is measured MB state vector, consisting of lateral and depth coordinates; $\mathbf{S}_k$ is the innovation covariance.

3. **Kalman motion model parameters**

Covariance of measurement noise **R**, covariance of model prediction noise **Q** and covariance of estimation noise P are parameters that need to be set for the linear Kalman motion model and sometimes can significantly reduce tracking accuracy with inappropriate values. **R** can be set by the localisation uncertainty of the imaging system, which can be measured by *in vitro* experiment and estimated for *in vivo* acquisition. Although **Q** can be chosen as a reasonable value based on the engineering practice, we proposed a method to estimate **Q** from data to avoid human interaction and reduce effort in tunning parameters when dealing with various vasculature and hemodynamics.

As presented in the last section, a linear motion model is used to predict MB movement in the third frame using the first two frames. Steps from 1 to 3 in the last section can be used for to estimate **Q** for each three adjacent frames. Instead of saving distance between the prediction and observation for each MB, the discrepancies along lateral $e^{x,m} = (x^q + x^m - 2x^n)$ and depth $e^{z,m} = (z^q + z^m - 2z^n)$ were saved for each MB in the first of the three frames. By assuming most of pairing after these three steps are right or close to the right, the discrepancies approximate the measure pre-fit residual of Kalman filtering for each MB movement in the time series. Therefore, we used the variance of the discrepancies calculate the pre-fit residual covariance.

$$\hat{\mathbf{S}} = \begin{bmatrix} \sigma_{ex}^2 & 0 \\ 0 & \sigma_{ez}^2 \end{bmatrix} \tag{12}$$

where $\sigma_{ex}^2$ and $\sigma_{ez}^2$ are the variance across $e^{x,m}$ and $e^{z,m}$ respectively. As variance estimation becomes more accurate with a larger number of data, discrepancies were obtained from ten groups of three consecutive frames picked from the image sequence, and a variance $\sigma_e^2$ was calculated with all the discrepancies along two directions and used for $\sigma_{ex}^2$ and $\sigma_{ez}^2$.

The **S** in one frame was defined by

$$S_k = HP_{k|k-1}H^T + R$$
$$P_{k|k-1} = FP_{k-1|k-1}F^T + Q \qquad (13)$$

where $P_{k-1|k-1}$ and $P_{k|k-1}$ are the updated and predicted estimation covariance matrices respectively. Using optimal Kalman gain, $P_{k-1|k-1}$ is independent from the measurements and can achieve an asymptotic value as

$$P_\infty = F\left(P_\infty - P_\infty H^T(HP_\infty H^T + R)^{-1}HP_\infty\right)F^T + Q \qquad (14)$$

Therefore, $P_\infty$ is determined by the defined **R** and **Q** and used for $P_{k-1|k-1}$ to make MB Kalman filter independent from how many frames the MB have appeared in the imaging plane. The time continuous noise model [2] was used and thus

$$Q = \begin{bmatrix} \frac{1}{3f^3} & \frac{1}{2f^2} & 0 & 0 \\ \frac{1}{2f^2} & \frac{1}{f} & 0 & 0 \\ 0 & 0 & \frac{1}{3f^3} & \frac{1}{2f^2} \\ 0 & 0 & \frac{1}{2f^2} & \frac{1}{f} \end{bmatrix} q^2 \qquad (15)$$

where $q$ is the variance of error when assuming constant velocity in the linear Kalman motion model. Then, estimating **Q** was replaced by estimating $q$. As analytical solution of Eq. (14) was not found, the dichotomy was used to find $q$ in Algorithm 1.

---
**Algorithm 1** Find $q$
---
**Require:** **R**, $|\hat{S}|$, $[q_b, q_u]$ searching range of $q$
   **while** $q_u - q_b > \epsilon$ **do**
      $q \leftarrow (q_b + q_u)/2$
      Calculate $P_\infty$ with **Q** and **R**
      Calculate **S** with Eq. (13)
      **if** $|S| > |\hat{S}|$ **then**                    ➢ $|S|$ increases with $q$ in the searching range
         $q_u \leftarrow q$
      **else**
         $q_b \leftarrow q$
      **end if**
   **end while**
**Return** $q$
---